\documentclass[]{emulateapj}

\usepackage{natbib}

\shorttitle{Multiple components of early-type dwarfs}
\shortauthors{Janz et al.}

\begin{document}

\submitted{}
\title{Dissecting early-type dwarf galaxies into their multiple components}

\author{J.~Janz\altaffilmark{1,2,*}, E.~Laurikainen\altaffilmark{3,1}, T.~Lisker\altaffilmark{2}, H.~Salo\altaffilmark{1}, R.~F.~Peletier\altaffilmark{4}, S.-M.~Niemi\altaffilmark{5}, M.~den~Brok\altaffilmark{4}, E.~Toloba\altaffilmark{6}, J.~Falc\'on-Barroso\altaffilmark{7,8}, A.~Boselli\altaffilmark{9}, G.~Hensler\altaffilmark{10}}
\email{jjanz@ari.uni-heidelberg.de}

\affil{
$^1${Division of Astronomy, Department of Physics, P.O. Box 3000,  FI-90014 University of Oulu, Finland}\\
$^2${Astronomisches Rechen-Institut, Zentrum f\"ur Astronomie der Universit\"at Heidelberg, M\"onchhofstra{\ss}e 12-14, D-69120 Heidelberg, Germany}\\
$^3${Finnish Centre for Astronomy with ESO (FINCA), University of Turku, Finland}\\
$^4${Kapteyn Astronomical Institute, University of Groningen, PO Box 800, 9700 AV Groningen, the Netherlands}\\
$^5${Department of Physics and Astronomy, University of North Carolina, Chapel Hill, CB 3255, Phillips Hall, Chapel Hill, NC 27599-3255, USA}\\
$^6${UCO/Lick Observatory, University of California, Santa Cruz, 1156, High street, Santa Cruz, CA 95064}\\
$^7${Instituto de Astrof\'isica de Canarias, V\'ia L\'actea s/n, La Laguna, Tenerife, Spain}\\
$^8${Departamento de Astrof\'isica, Universidad de La Laguna, E-38205 La Laguna, Tenerife, Spain}\\
$^9${Laboratoire d'Astrophysique de Marseille, UMR 6110 CNRS, 36 rue F.~Joliot-Curie, F-13388 Marseille, France}\\
$^{10}${University of Vienna, Institute of Astronomy, T\"urkenschanzstra\ss e 17, 1180 Vienna, Austria}
}











\altaffiltext{*}{Fellow of the Gottlieb Daimler and Karl Benz Foundation.}

\begin{abstract}
 Early-type
dwarf galaxies, once believed to be simple systems, have recently been shown to exhibit an intriguing
diversity in structure and stellar content. 
To analyze this further, we started the \textit{SMAKCED} project\footnote{$S$tellar content, $MA$ss and $K$inematics of $C$luster $E$arly-type
 $D$warfs, \url{http://www.smakced.net}}, and obtained deep $H$-band images for 101 early-type dwarf galaxies
in the Virgo cluster in a brightness range of $-19\le M_r\le-16$ mag, 
typically reaching a signal-to-noise of 
1 per pixel of $\sim0.25\arcsec$ at surface brightnesses $\sim22.5$ mag/arcsec$^2$ in the $H$-band. 
Here we present the first results 
of  decomposing  their 
two-dimensional light distributions. 
This is the first study dedicated to early-type dwarf galaxies using the two-dimensional multi-component
decomposition approach, which has been proven to be important for giant galaxies.
Armed with this new technique, we find more structural components than previous studies:
only a quarter  of the galaxies fall into the simplest group, namely those represented by a single S\'ersic function,
optionally with a nucleus. 
Furthermore, we find 
 a bar fraction of 18\%. 
We detect also a similar
 fraction of lenses which appear as shallow structures with sharp
 outer edges. Galaxies with bars and  lenses are found to be more concentrated
 towards the Virgo galaxy center than the other sample galaxies. 
 \end{abstract}


\section{Introduction}
Early-type dwarf (dE) galaxies are the most abundant galaxy population in
high-density environments. Their low mass and large number make them 
ideal probes of the mechanisms that can alter the appearance of
galaxies: internal processes as well as environmental influences.
Their ubiquity  and susceptibility to various physical mechanisms give them a
 key role in understanding galaxy cluster evolution. 
The popular belief that dEs were formed from spiral and irregular galaxies
at late epochs by the cluster environment (e.g.~\citealt{1998ApJ...495..139M,Boselli,KormendyBender,Toloba:2011p3986}) is
contrasted with the formation of dEs in models of a
$\Lambda$CDM universe, as the descendants of cosmological building
blocks. In the latter scenario dEs would be close relatives to
their giant counterparts (e.g.~\citealt{deRijcke:2005p63,Janz:2008p151,Janz:2009p780,Weinmann:2011p3976}).

Disk structures in dEs
have been searched for since the early 1990s \citep{binggeli_cameron,Jerjen:2000p1547,2002A&A...391..823B}.
The prevailing interpretation was that the disk structures, as imprints of their
host galaxy's history, point at late-type disk galaxies as progenitors for dEs.
Subsequent work in recent years has shown that dEs are rather heterogeneous, with
 their various characteristics depending strongly on the position within the cluster: different
subclasses based on morphology and stellar population characteristics 
were identified \citep{Lisker:2006p385,Lisker:2006p392,Lisker:2007p373,Lisker:2008p372}; 
dEs are not old in general, but cover a large range in age
and metallicity  \citep{Michielsen:2008p3941,Paudel:2010p4315}; the
degree of rotational support varies \citep{Toloba:2009p3937,Toloba:2011p3986}; and
their shapes seem  to depend even on their orbital characteristics \citep{Lisker:2009p3975}.
The various results imply that the dEs might be a mixed bag,  possibly with multiple formation channels \citep{LiskerAN}.

Here, we explore the feasibility of revealing detailed structures in dEs using a
two-dimensional multi-component decomposition approach.
\citet{Laurikainen:2010p4602} demonstrated for S0s 
how morphological information
via such detailed  decompositions 
can contribute to unveiling possible formation processes of those galaxies.
Surprisingly, we find for  the dEs that only a minority of  them in the Virgo cluster appears  to follow the classical picture of a featureless galaxy.


\section{Data}
\begin{figure}
\epsscale{1.2}
\plotone{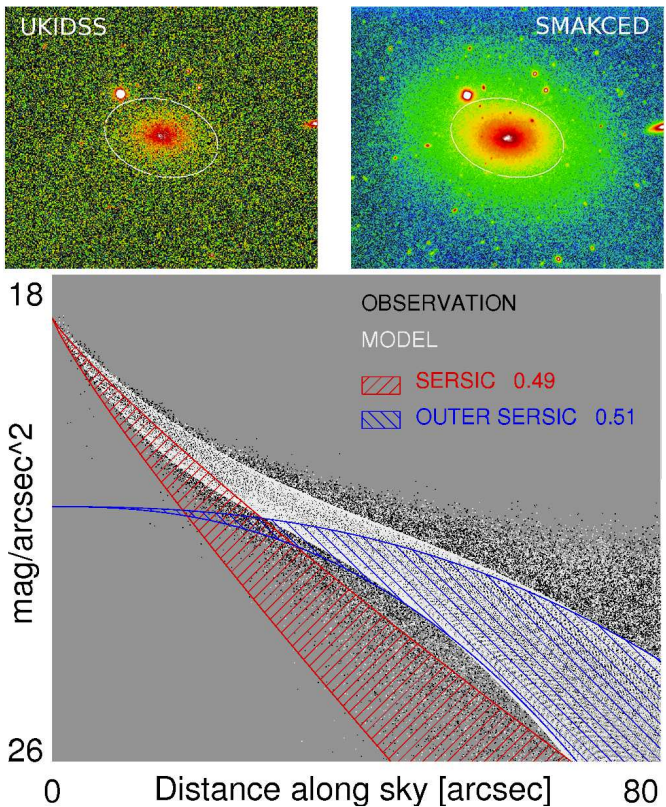}
\caption{One example (VCC0170) of the observed two-dimensional light profiles with the fitted components. The observation (every unmasked pixel in the image) is displayed with black points, the total model in white and the individual components with shaded areas. The numbers quote the fractions of light in the components. For a direct comparison with the observation the model is plotted twice: the pure model is the white band with well defined boundaries. For another illustration of the model (white dots spread among the black dots of the observation) random values were  added according to the noise in the observed image. For clarity, the nuclear component, which is slightly offset in this galaxy, is not plotted. Top panels show the UKIDSS NIR image \textit{(left)} and our image obtained with ESO NTT \textit{(right)}. The optical half-light radius ($r_e=32\arcsec$, \citealt{Janz:2008p151}) is indicated with an ellipse. North is left.}
\label{fig:profile_plot}
\end{figure}

Within the \textit{SMAKCED} project$^a$  we aim at obtaining deep $H$-band images for a complete sample of 174
 early-type galaxies in the Virgo cluster in the brightness range of $-19\le M_r\le -16$ mag ($[m-M]_{\textit{Virgo}}=31.09$ mag, \citealt{Mei:2007p872}). 
  The sample is drawn from the Virgo cluster catalog (VCC, \citealt{Binggeli:1985p849}, \textit{`certain members'}).
Observations in the near-infrared (NIR) allow the most direct characterization
 of the galaxies' stellar mass distribution and are much less affected by dust extinction than in the optical. Some dEs contain dust \citep{Peletier93,Lisker:2006p392,dust}, which could mimic complex structures.
 During 2010-2011
 we obtained images for 81 of the 174 galaxies with ESO NTT, TNG, and NOT.  
 Complemented with archival data this yields a sample of 101 galaxies. Our images typically reach a $H$-band surface brightness of 22.2-23.0 mag/arcsec$^2$
 at $S/N=1$ per pixel (scale $0.234\arcsec$ to $0.288\arcsec$),  deeper than in previous works (Fig.~\ref{fig:profile_plot}). 
The reductions of the on-target dithered observations, done with \textsc{iraf}, included cross-talk removal,
flatfielding, sky subtraction, and correction for the field distortions, where necessary.

\setcounter{footnote}{0}
\section{Two-dimensional Decompositions}
For decomposing the galaxies' light into potentially multiple components we employ \textsc{galfit} 3.0 \citep{Peng:2010p4252}, which
uses a $\chi^2$ minimization algorithm to find the optimal solution for a given set of functions and starting values. \textsc{galfit}'s solutions were visually  evaluated using \textsc{galfidl}\footnote{H.~Salo,  \url{http://sun3.oulu.fi/~hsalo/galfidl.html}}, a set of \textsc{idl} routines. Fore- and background sources were masked out;  the PSF FWHM was determined for several point-like sources using \textsc{SExtractor} on each image (typically $\sim0.9\arcsec$) 
and the model was convolved with the averaged (gaussian) PSF  during the fitting process. 

\begin{figure*}
\includegraphics[scale=0.71,angle=-90]{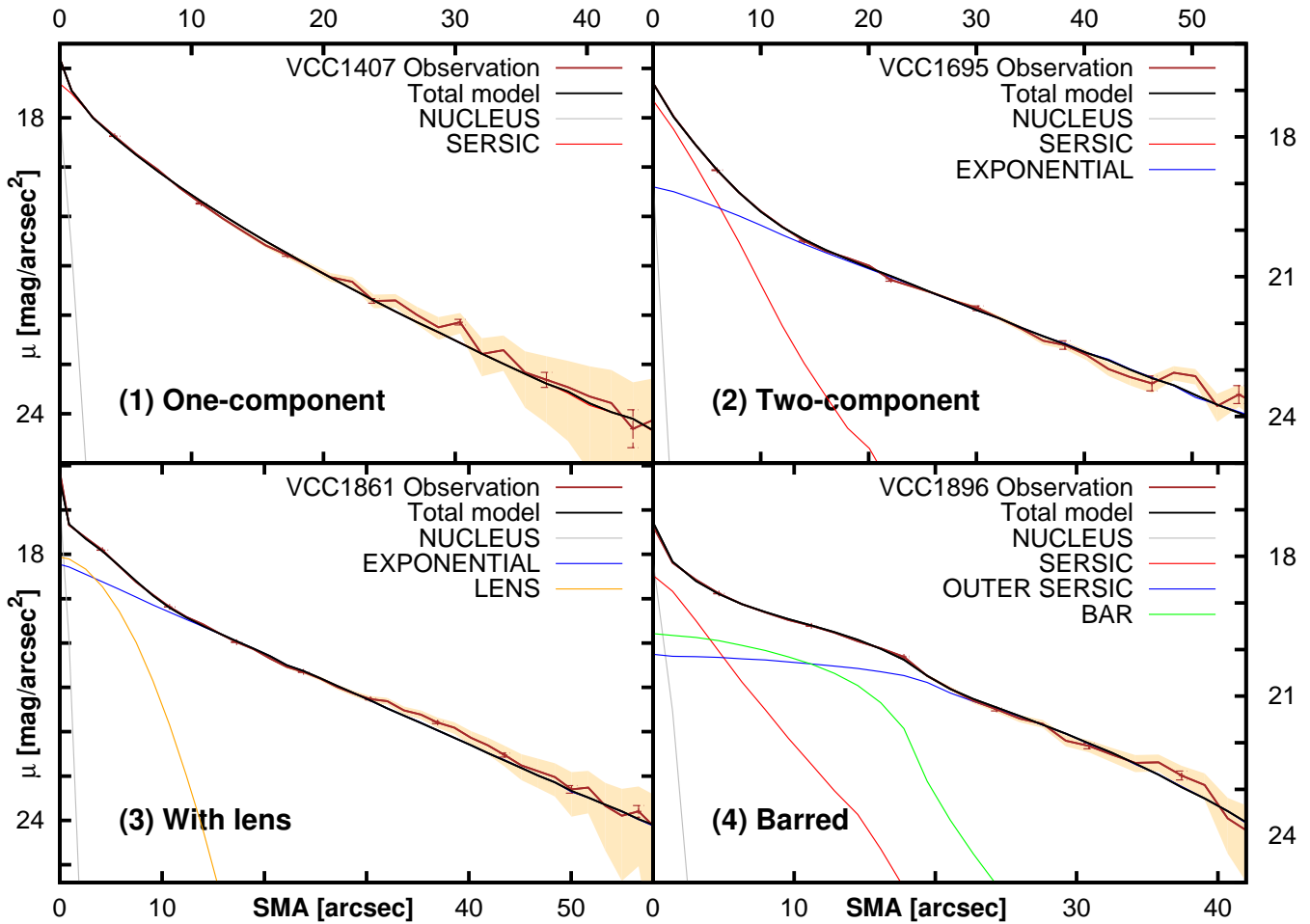}
\caption{Groups of structural types. We identified four characteristic, distinct structural types by decomposing the two-dimensional light distributions of the galaxies  (see Sec.~\ref{sec:groups}). The panels show  light profiles for representative galaxies  (SMA = semi-major axis). The shaded areas display the maximal systematic error, estimated by adding and subtracting the large-scale background variations' RMS to the intensities. The error bars indicate the intensity uncertainties as measured by \textsc{iraf}/\texttt{ellipse}.}
\label{fig:profile_types}
\end{figure*}

As further input \textsc{galfit} needs the uncertainties of the pixel values, i.e.~a $\sigma$-image.
We  calculate, for each pixel, the standard deviation over the individual images,  taking advantage of the large number of  exposures for each galaxy. 
The obtained  $\sigma$-image is normalized to correspond to the sky pixel-to-pixel variations in the final coadded image.
The model fitting does not incorporate possible systematic large scale background variations. The potential bias by such variations is estimated from  the RMS of mean sky values in small boxes distributed on the image. Such a bias could alter the profiles within the shaded area indicated in Fig.~\ref{fig:profile_types}.

For all galaxies we fitted the following basic models: one-component, one-component+nucleus, and two-component models. \citet{1963BAAA....6...41S} functions were used for the components, while the nucleus was modeled by a point source. Subsequently, we visually evaluated the quality of the fit by inspecting the residual structures seen in the model-subtracted images (Fig. 3) and in profile representations showing all observed pixels (Fig. 1). We also inspected the one-dimensional surface brightness profiles and profiles of position angle and ellipticity (obtained by  \textsc{iraf}/\texttt{ellipse} fitting). When deemed necessary from the fit residuals and profile representations, we fitted models with additional bar or lens components. 
Lens, in this context, refers to an inner component with `a shallow brightness gradient with a sharp outer edge' \citep{Kormendy1979}. 
In distinction to bars ($b/a\lesssim0.5$), lenses have intrinsic axial ratios close to unity ($b/a>0.7$).

The basic models consist of one or two S\'ersic functions. The fit of the outer component in the two-component model had a S\'ersic index $n$ fixed to 1 (i.e. exponential). Its orientation and ellipticity were fixed to the mean of the outer isophotes. For a few galaxies $n$ was a free parameter, to account for a steeper drop of the outer profile with $n < 1$ (Fig.\ \ref{fig:profile_plot} and panel 4 of Fig.\ \ref{fig:profile_types}). 
For a substantial number of the two-component galaxies, the S\'ersic index of the inner component is $n\approx1$ ($n < 1.2$ for 53\%). For bars and lenses we chose Ferrers' function (see \citealt{Peng:2010p4252}) with elliptical isophotes, since it allows for a better treatment of the outer cutoff of the surface brightness (see \citealt{Laurikainen:2009p206}). In a few cases the fit was considerably improved when  the center for each component was left free. Notably most of those galaxies were classified as having residual star formation in the center (dE(bc)s,  \citealt{Lisker:2006p392}).

\section{Results}
\subsection{Groups of structural types}
\label{sec:groups}
\begin{figure}
\includegraphics[scale=0.54]{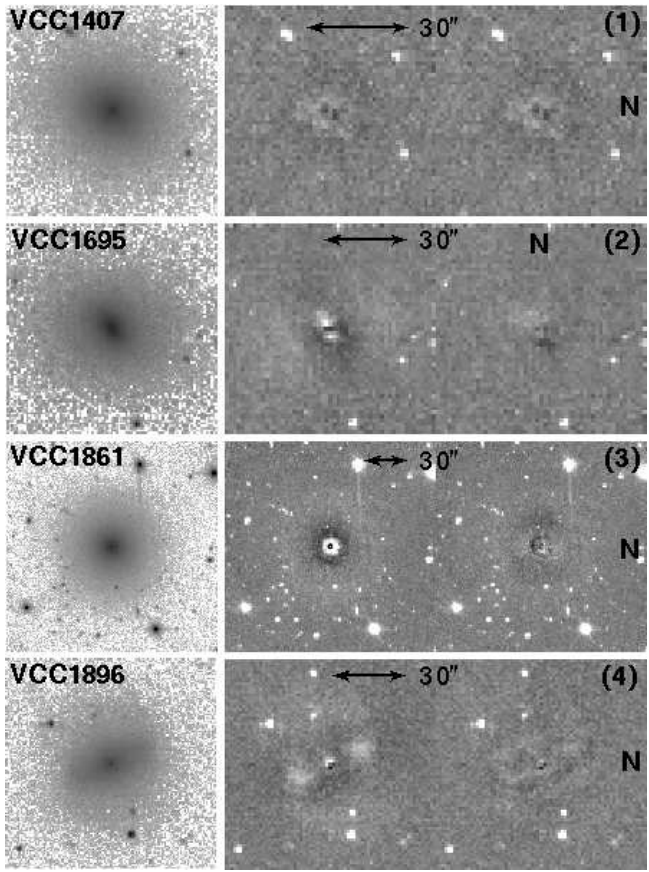}\\
\caption{Comparison of simple and final models for the galaxies in Fig.~\ref{fig:profile_types}. In each row a cutout of the image is shown to the left. The residuals after subtracting  a simple S\'ersic+nucleus \textit{(middle)} and the final model \textit{(right)} show the improvement. The grayscales show $\pm 3\sigma$ for the residual images, north is the direction indicated by \emph{N}. }
\label{fig:residuals}
\end{figure}

We order the galaxies by  the set of components building up the galaxy's model
and define the following groups: \emph{(1)} one component, \emph{(2)}   two components (typically
 S\'ersic+exponential), \emph{(3)} galaxies with lenses, and \emph{(4)} barred galaxies. Representative examples for the groups 
are shown in Fig.~\ref{fig:profile_types}. 
The final model for a given galaxy, and thus its group assignment, was chosen based on the visually judged improvement of the residual structures (Fig.~\ref{fig:residuals}) and profile representations.

In most cases the  lens is accompanied by an exponential outer component;
however in two cases it is a S\'ersic component with $n>1$. Only  3 of the galaxies with a lens  
have three components (not counting the nucleus).
The galaxies with lenses might be regarded as  two-component systems with further complexity  (cf.\ \citealt{binggeli_cameron}), but we  assign them to their own group. 
The presence of a nucleus does not affect the group designation. For in-depth analysis
of nuclei in dEs we refer to the literature, especially to studies resolving them (i.e. Advanced Camera for Surveys (ACS) Virgo Cluster
Survey,  \citealt{Ferrarese:2006p586}).
Spiral arms can be clearly seen in the residual image of four galaxies, while for two more galaxies  they are at the detection limit. Two of the galaxies with a lens host additionally a small bar.

It is not possible  to assign all galaxies  unambiguously to one group, since two different models might fit equally well. Uncertainties that would shift the galaxies from the two- to  one-component group include:
\emph{(a1)}  the improvement of two components over the one-component model is marginal, \emph{(a2)} the inner component of a galaxy fitted by a S\'ersic function  is so small that it might be just 
 a nuclear component. If the edge of a lens is less well-defined, the
 component can be alternatively fitted  with a S\'ersic function with $n<1$, which would shift the galaxy  from
 the group with lenses to the two-component group \emph{(b)}. Bars are characterized by high ellipticity and possibly an orientation distinct from 
 the disk component, but they are modeled with the same function as  lenses. Especially in more inclined galaxies
 this distinction can become difficult, the less certain cases being  listed as \emph{(c)}.
  
 In Table~\ref{tab:groups} we summarize the number statistics for the four groups and number the less certain cases among them.
The sample is divided into  different bins of galaxy brightness, morphological subclass, and projected clustercentric distance.\footnote{We consider the cluster center to be marked by M87.} 
In total  22 galaxies are not included in any group as their decomposition is unreliable:
14 are more inclined than $65^\circ$, i.e.\ axis ratio $<\cos(65^\circ$),  and for  8 galaxies
 no satisfactory fit was achieved due to persistent residuals structures.
 ~\\
\subsection{Analysis}
First and foremost our analysis separates galaxies that satisfy the simplest one-component models from those with more complex structures. 
The fraction of simple galaxies is surprisingly low, given the picture of dEs as structureless, red and dead 
galaxies: only 24\% of the galaxies in our analyzed sample exhibit a simple structure (see Table 1, 27\% of those with dE classification;  34\% when adding the less certain two-component galaxies). 
Obviously, our low fraction of simple galaxies  might even decrease if still deeper images were available. 

The largest fractions of simple galaxies are found among 
those where no disk feature or blue center had been identified previously \citep{Lisker:2007p373}, and among the faintest galaxies.
That  the fraction of galaxies with multi-component structures 
 increases towards brighter galaxies in Table~\ref{tab:groups} is not a simple selection effect, since the desired image depth was chosen according to the surface brightness at the half-light radius of each galaxy. 
 Therefore, we do not expect to miss components  in dimmer galaxies provided that they contain a similar fraction of light and have similar relative extent.
 While galaxies of higher mass may be expected to shield their structures more easily against external heating, 
the increase of the multi-component galaxy fraction with increasing galaxy brightness
is not statistically significant: 
 a Kolmogorov-Smirnov (KS)-test yields 7.2\% probability for them being drawn from the same brightness
 distribution.

\begin{figure}
\includegraphics[scale=0.49]{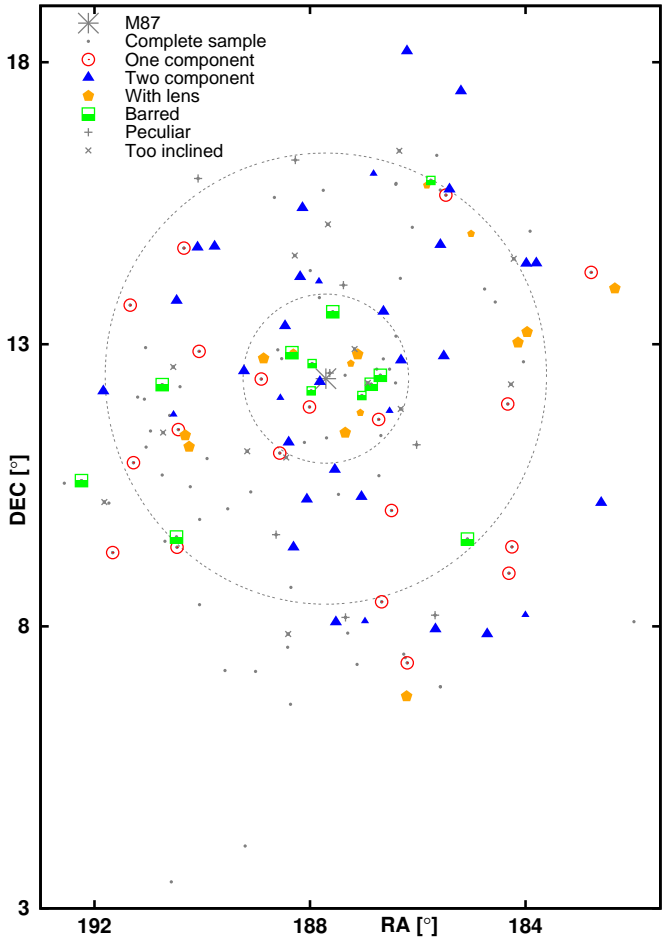}
\caption{Distribution of the galaxies in different groups inside the Virgo cluster.
Galaxies with less certain decompositions (see  Sec.~\ref{sec:groups} and Table 1) and galaxies that were not analyzed  are shown with smaller symbols.
 The circles indicate the radial distances of $1.5^\circ$ and $4^\circ$ from M87. }
\label{fig:cluster}
\end{figure}

Fig.~\ref{fig:cluster} displays the distribution of galaxies in the different groups inside the Virgo cluster.
Simple galaxies appear less centrally concentrated (see also Table~\ref{tab:groups}), but the
clustercentric distances of simple and multicomponent galaxies
do not differ significantly according to a KS-test. Indeed, simple
galaxies are distributed similarly to two-component galaxies, which
constitute the major part of the multicomponent ones. On the other hand,
barred galaxies are more concentrated towards the projected cluster center than 
the two-component galaxies (KS-test 1.6\%; 4.6\% when comparing barred and one-component galaxies). 

Bars were fitted in 12 galaxies, 4 of them being classified as less certain. With the two small bars in  galaxies \emph{with lenses}, not fitted in the decompositions,
this leads to a bar fraction  of 18\%  (13\% without less certain cases). Some galaxies in our sample fall into the dwarf brightness regime, but were previously
classified as E or S0. Omitting these galaxies does not change the bar fraction.
Lenses were fitted in 14 galaxies (18\%). 
\citet{Kormendy1979} suggested that bars and lenses are evolutionary related, i.e.\ that lenses are dissolved bars. When we treat barred galaxies and galaxies with a lens as one combined group, we find their projected clustercentric distance distribution to be significantly different from that of the other dEs (KS-test 0.4\%).

\begin{deluxetable*}{lrrrrrr}
\tablecaption{Frequencies of structural types}
\tabletypesize{\footnotesize}
\tablenum{1}
\tablecolumns{7}
\tablehead{
\colhead{Group} & \colhead{1} & \colhead{2 } & \colhead{3} & \colhead{4} & \colhead{Total}& \colhead{Complete}\\
 \colhead{} & \colhead{1 component} & \colhead{2 component} & \colhead{w. lens} & \colhead{barred} & \colhead{analyzed}& \colhead{sample}\\ 
  \colhead{Less certain cases} & \colhead{} & \colhead{(a1+a2)} & \colhead{ (b)} & \colhead{ (c)} & \colhead{}& \colhead{} }
\startdata
 
  All & 19 / 24\%  &  34 (7)  / 43\% &  14 (5)  / 18\% &  12 (4) /  15\%   & 79 & 174\\
\tableline
$-19\le M_r\le-18$ &1   / \, 8\%  & 5 (0)   / 38\%  & 4 (3)  /  31\%  & 3 (0)  /  23\% & 13 & 28\\
$-18<M_r\le-17$ & 10  / 21\%  &  23 (5) /  49\% &  7 (1)  / 15\%  &  7 (4) /  15\%  & 47 & 61\\
$-17<M_r\le-16$ & 8  / 42\%  & 6 (2) /   32\%  & 3 (1)   / 16\%  & 2 (0)  /  11\% & 19 & 85\\
  \tableline
dE(all) & 18  / 27\% & 29 (5)  / 43\%& 10 (4) /  15\%&10 (4)  / 15\%  & 67Ê& 145\\
\tableline
dE(N) & 11  / 37\%  &  8 (1) /  27\%  &  8 (3)  / 27\%  &  3 (2)  / 10\%  & 30 & 106 \\
dE(nN) & 2  / 40\%  & 3 (0) /   60\%  & 0   & 0   &   5 & 27\\
dE(di) & 4  / 16\%  & 11 (2)   / 44\%  & 2 (1)  /  \, 8\%  & 8 (2)  /  32\%  & 25 & 33 \\
dE(bc) & 1 /  10\%  & 9 (2)  / 90\%  &  0 &  0  &  10 & 15 \\
E \& S0 & 1 /  \, 8\% &  5 (2)  / 42\%  & 4 (1)  /  33\% & 2 (0)  / 17\%  & 12 & 29\\
\tableline
$D_{M87}<1.5^\circ$ & 3  / 13\%  & 7 (2)  /  30\%  & 6 (3)  / 26\%  &  7 (3)  / 30\% & 23 & 42 \\
$1.5^\circ\le D_{M87}<4^\circ$ & 9  / 26\%  &15 (3)  /  44\%  & 6 (2) /  18\%  &  4 (1)  / 12\%  & 34 & 84\\
$D_{M87}\ge4^\circ$ & 7  / 32\%  & 12 (2)  /  55\%  & 2 (0)  /  \,  9\%  &  1 (0)  / \, 5\%   & 22 & 48
\enddata
\tablecomments{We list numbers and fractions of galaxies in each group  binned over the brightness,  $dE$ subclass \citep{Lisker:2006p385,Lisker:2006p392,Lisker:2007p373}, and  angular distance to M87  ($1^\circ = 0.284$ Mpc). The total sum in the subclass binning is larger than the total number of galaxies, since $dE$s can belong to  multiple subclasses.
The less certain cases in each group ($a1$, $a2$, $b$, and $c$, see Sec.~\ref{sec:groups}), shown in parenthesis, are included in the numbers and fractions. Additionally two small bars were visually identified but not fitted and therefore not counted in this table.\label{tab:groups}}
\end{deluxetable*}

\section{Discussion}
Recent studies of detailed structures in the dEs in the Virgo cluster include \citet{Mcdonald:2011p4445}, \citet{Ferrarese:2006p586}, and \citet{Lisker:2006p385}. 
\citet{Mcdonald:2011p4445} fitted S\'ersic and S\'ersic+exponential models to the one-dimensional
light profiles of galaxies in the Virgo cluster, also in the $H$-band. 
Number statistics of structural components were not given, but taking their decompositions for our sample leads to a very similar fraction of simple galaxies. The agreement
in a one-by-one comparison for the galaxies in common is less good, though.
A fair comparison for individual cases is hindered, since the algorithm by
which their fitting code decides in favor of one or two components has not been described in detail.

Ferrarese et al.~focused more on the inner regions with the superior 
resolution of HST ACS.
\citet{Lisker:2006p385} searched  systematically for disk features using Sloan Digital Sky Survey images. They
introduced the dE subclass named dE(di) for galaxies, in which such signatures were revealed by unsharp masking.
The brightness distribution of the fraction of our galaxies fitted by more than one component is similar to their fraction of galaxies with disk features:  up to 50\%  for the brightest dEs, but decreasing
towards fainter brightnesses. 
Concerning morphological types, we find that 84\% of the 25 dE(di)s in our sample have multiple components.
In eight of them we fit a bar.  Three of the four dE(di)s fitted with only one component show spiral arms  in the residual images.
Also, 64\% of the dE(N)s  in our sample (in their terminology, nucleated dEs with no other feature)
show more complex morphologies.
Already \citet{binggeli_cameron} noticed visually that two thirds of dEs in our magnitude range show a break in their B-band surface brightness profile, but recent studies did not take this up.

\citet{2005AJ....130..475A}  decomposed  the azimuthally averaged light profiles for 99 dEs  in the Coma cluster,  in a magnitude range of 
$-16\le M_B\le-18$ mag.
Their criterion for the need of two components (S\'ersic+exponential) was a deviation of more than 0.15 mag 
of the simple S\'ersic model from the observed profile at any radius, taking into account photometric errors. 
They found that 34\% of their galaxies with reliable photometry are not well fitted by a single
S\'ersic function. Our fraction for the Virgo cluster is much larger, 82\%, for a comparable magnitude range of $-17\le M_r\le-19$ mag.
It would be interesting whether the difference is due to their simpler method and worse physical
resolution, or due to a real difference,
for example caused by the environment. 
 \citet{Hoyos:2011p4620} fitted single S\'ersic profiles for  galaxies in the Coma cluster region using \textsc{galfit} 
and \textsc{gim2d} on HST ACS data. Since their
physical resolution is comparable to our study,  their planned multi-component decompositions will provide an interesting comparison sample. 

Even though Coma is more massive, denser and has a larger fraction of red galaxies than Virgo (\citealt{Weinmann:2011p3976}, and references therein),
the bar fraction we find in Virgo (and its decrease towards fainter galaxy brightnesses, see Table 1) agrees with the 
findings of \citet{MA} for the Coma cluster.
One may speculate that the increase of the Virgo cluster bar fraction towards the dense cluster center, and yet the similarity of the overall Virgo bar fraction to the denser Coma cluster is due to an interplay between tidal interactions that induce bar formation and heating of the disks that impedes the formation and longevity of bars. 
%
%
Also bars in spiral galaxies in the Virgo cluster \citep{Andersen}, in disk galaxies in the Coma cluster \citep{Thompson} and in clusters at intermediate redshifts \citep{Barazza09} were found to be more frequent towards the cluster centers.


While unsharp masking is sensitive to sharper features  like bars, parametric functions fitted to the one-dimensional light profile are sensitive to deviations from a simple form, i.e. the S\'ersic profile shape.
Our two-dimensional fitting technique accounts for both of them in a quantitative manner.
In this sense our smaller fractions of simple galaxies as compared to \citet{Lisker:2006p385}, \citet{2005AJ....130..475A}, and earlier
studies may well be  a consequence of the different methods used.

\citet{Paudel:2010p4315} analyzed ages and metallicities of a sample of Virgo dEs using Lick indices. They found that the dE(di)s 
have younger stellar populations on average than the dE(N)s (see also \citealt{Toloba:2009p3937}), but that also the brighter dEs, for which disk features are more frequent,
have younger populations than the fainter ones. They speculate that further disks may be present in the dE(N)s but may 
have eluded discovery so far, which is interesting given our large fraction of galaxies with complex profiles. 
If late-type galaxies had been transformed into dEs \citep{KormendyBender,Boselli}, the complex structures could be understood as inherited from the progenitor. At least part of the late-type galaxies with stellar masses between $\sim10^9$ and $\sim10^{10}\,{\rm M_\odot}$ possess a two component structure \citep{GrahamWorley}. If this structure survived disk thickening, caused by tidal heating \citep{Gnedin,Smith10} and by gas depletion from ram-pressure stripping \citep{Smith11}, such galaxies may resemble complex dEs today.

\section{Summary}
For the first time, we have applied a detailed two-dimensional multi-component fitting technique 
to a large sample of dEs, using deep NIR images. The images typically reach surface brightnesses of 22.2-23.0 mag/arsec$^2$ in the $H$-band.
In many galaxies this method has revealed  more complex structures than previously known.
Only for 27\% of the dEs is  the light distribution  well represented by a single S\'ersic function. 
Of the dE(di)s 16\% were fitted by one component, but most of those  appear to have disks manifested as spiral arms in the residual images. Bars were detected in 18\% and lenses also in 18\% of the galaxies.
The physical nature of the various components remains to be investigated, ideally with a combined kinematical and stellar population analysis in future studies.

\acknowledgments
We thank H.T.~Meyer, K.S.A.~Hansson, and S.~Paudel for helping with the observations.
We  thank Magnus G\aa lfalk 
and Amanda Djupvik 
for their kind help related to the NOT.
TL~thanks Chris Flynn and Burkhard Fuchs for initiating collaborations.

JJ~acknowledges support by the Gottlieb Daimler and Karl Benz Foundation.
TL~and JJ~are supported by the German Research Foundation (DFG, GSC 129/1). 
HS and JJ are supported by the Academy of Finland.
JFB acknowledges support from the Ram\'on y Cajal Program and from grant AYA2010-21322-C03-02 by the Spanish Ministry of Science and Innovation.

Observations were collected at the Nordic Optical Telescope, the Telescopio Nazionale Galileo, and the European Organisation for Astronomical Research in the Southern Hemisphere (064.N-0288, 085.B-0919), based on proposals by the SMAKCED team (\url{http://www.smakced.net}).


\begin{thebibliography}{23}
\expandafter\ifx\csname natexlab\endcsname\relax\def\natexlab#1{#1}\fi

\bibitem[{Aguerri {et~al.}(2005)Aguerri, Iglesias-P{\'a}ramo, V\'ilchez,
  Mu{\~n}oz-Tu{\~n}{\'o}n, \& S{\'a}nchez-Janssen}]{2005AJ....130..475A}
Aguerri, J.~A.~L., Iglesias-P{\'a}ramo, J., V\'ilchez, J.~M.,
  Mu{\~n}oz-Tu{\~n}{\'o}n, C., \& S{\'a}nchez-Janssen, R. 2005, \aj, 130, 475
  
  \bibitem[{Andersen (1996)}]{Andersen}
  Andersen, V. 1996, AJ, 111, 1805
  

\bibitem[{Barazza {et~al.}(2002)Barazza, Binggeli, \&
  Jerjen}]{2002A&A...391..823B}
Barazza, F.~D., Binggeli, B., \& Jerjen, H. 2002, \aap, 391, 823

  \bibitem[{Barazza {et~al.}(2009)Barazza et al.}]{Barazza09}
Barazza, F.~D., {et~al.} 2009, A\&A, 497, 713
  

\bibitem[{Binggeli {et~al.}(1985)Binggeli, Sandage, \&
  Tammann}]{Binggeli:1985p849}
Binggeli, B., Sandage, A., \& Tammann, G.~A. 1985, AJ, 90, 1681

\bibitem[{Binggeli \& Cameron(1991)}]{binggeli_cameron}
Binggeli, B., \& Cameron, L.~M. 1991, \aap, 252, 27

\bibitem[{Boselli {et~al.}(2008)Boselli, Boissier, Cortese, \&
  Gavazzi}]{Boselli}
Boselli, A., Boissier, S., Cortese, L., \& Gavazzi, G. 2008, ApJ, 674, 742

\bibitem[{de~Looze {et~al.}(2010)de~Looze et~al.}]{dust}
de~Looze, I. et~al. 2010, A{\&}A, 518, 54

\bibitem[{de~Rijcke {et~al.}(2005)de~Rijcke, Michielsen, Dejonghe, Zeilinger,
  \& Hau}]{deRijcke:2005p63}
de~Rijcke, S., Michielsen, D., Dejonghe, H., Zeilinger, W.~W., \& Hau, G. K.~T.
  2005, A{\&}A, 438, 491
  

\bibitem[{Ferrarese {et~al.}(2006)Ferrarese, C{\^o}t{\'e}, Jord{\'a}n, Peng,
  Blakeslee, Piatek, Mei, Merritt, Milosavljevi{\'c}, Tonry, \&
  West}]{Ferrarese:2006p586}
Ferrarese, L., {et~al.} 2006, ApJS,  164,  334

\bibitem[{Graham \& Worley (2008)}]{GrahamWorley}
Graham, A.~W., Worley, C.~C., 2008, MNRAS, 388, 1708

\bibitem[{Gnedin (2003)Gnedin}]{Gnedin}
Gnedin, O.~Y. 2003, ApJ, 589, 752

\bibitem[{Hoyos {et~al.}(2011)Hoyos, den Brok, Kleijn, Carter, Balcells,
  Guzm{\'a}n, Peletier, Ferguson, Goudfrooij, Graham, Hammer, Karick, Lucey,
  Matkovi{\'c}, Merritt, Mouhcine, \& Valentijn}]{Hoyos:2011p4620}
Hoyos, C., {et~al.} 2011, MNRAS,  411, 2439

\bibitem[{Janz \& Lisker(2008)}]{Janz:2008p151}
Janz, J., \& Lisker, T. 2008, ApJ, 689, L25

\bibitem[{Janz \& Lisker(2009)}]{Janz:2009p780}
---. 2009, ApJ, 696, L102

\bibitem[{Jerjen {et~al.}(2000)Jerjen, Kalnajs, \& Binggeli}]{Jerjen:2000p1547}
Jerjen, H., Kalnajs, A., \& Binggeli, B. 2000, A{\&}A, 358, 845

\bibitem[{Kormendy(1979)}]{Kormendy1979}
Kormendy, J. 1979, ApJ, 227, 714

\bibitem[{Kormendy \& Bender(2011)}]{KormendyBender}
Kormendy, J., \& Bender, R. 2011, ApJS accepted,  arXiv:1110.4384


\bibitem[{Laurikainen {et~al.}(2009)Laurikainen, Salo, Buta, \&
  Knapen}]{Laurikainen:2009p206}
Laurikainen, E., Salo, H., Buta, R., \& Knapen, J.~H. 2009, ApJ, 692, L34

\bibitem[{Laurikainen {et~al.}(2010)Laurikainen, Salo, Buta, Knapen, \&
  Comer{\'o}n}]{Laurikainen:2010p4602}
Laurikainen, E., Salo, H., Buta, R., Knapen, J.~H., \& Comer{\'o}n, S. 2010,
  MNRAS, 405, 1089

\bibitem[{Lisker {et~al.}(2006{\natexlab{a}})Lisker, Glatt, Westera, \&
  Grebel}]{Lisker:2006p385}
Lisker, T., Glatt, K., Westera, P., \& Grebel, E.~K. 2006{\natexlab{a}}, AJ,
  132, 2432

\bibitem[{Lisker {et~al.}(2006{\natexlab{b}})Lisker, Grebel, \&
  Binggeli}]{Lisker:2006p392}
Lisker, T., Grebel, E.~K., \& Binggeli, B. 2006{\natexlab{b}}, AJ, 132, 497

\bibitem[{Lisker {et~al.}(2008)Lisker, Grebel, \& Binggeli}]{Lisker:2008p372}
---. 2008, AJ, 135, 380

\bibitem[{Lisker {et~al.}(2007)Lisker, Grebel, Binggeli, \&
  Glatt}]{Lisker:2007p373}
Lisker, T., Grebel, E.~K., Binggeli, B., \& Glatt, K. 2007, ApJ, 660, 1186

\bibitem[{Lisker(2009)}]{LiskerAN}
Lisker, T. 2009, AN 330, 1043

\bibitem[{Lisker {et~al.}(2009)Lisker, Janz, Hensler, Kim, Rey, Weinmann,
  Mastropietro, Hielscher, Paudel, \& Kotulla}]{Lisker:2009p3975}
Lisker, T., {et~al.} 2009, ApJ, 706, L124

\bibitem[{McDonald {et~al.}(2011)McDonald, Courteau, Tully, \&
  Roediger}]{Mcdonald:2011p4445}
McDonald, M., Courteau, S., Tully, R.~B., \& Roediger, J. 2011, MNRAS, 414,
  2055

\bibitem[{Mei {et~al.}(2007)Mei, Blakeslee, C{\^o}t{\'e}, Tonry, West,
  Ferrarese, Jord{\'a}n, Peng, Anthony, \& Merritt}]{Mei:2007p872}
Mei, S., {et~al.} 2007, ApJ, 655, 144

\bibitem[{M\'endez-Abreu {et~al.}(2010)M\'endez-Abreu, S\'anchez-Janssen, \& Aguerri }]{MA}
M\'endez-Abreu, J.; S\'anchez-Janssen, R.; Aguerri, J. A. L. 2010,
ApJ, 711, L61

\bibitem[{Michielsen {et~al.}(2008)Michielsen, Boselli, Conselice, Toloba,
  Whiley, Arag{\'o}n-Salamanca, Balcells, Cardiel, Cenarro, Gorgas, Peletier,
  \& Vazdekis}]{Michielsen:2008p3941}
Michielsen, D., {et~al.} 2008, MNRAS, 385, 1374

\bibitem[{Moore {et~al.}(1998)Moore, Lake, \& Katz}]{1998ApJ...495..139M}
Moore, B., Lake, G., \& Katz, N. 1998, \apj, 495, 139

\bibitem[{Paudel {et~al.}(2010)Paudel, Lisker, Kuntschner, Grebel, \&
  Glatt}]{Paudel:2010p4315}
Paudel, S., Lisker, T., Kuntschner, H., Grebel, E.~K., \& Glatt, K. 2010,
  MNRAS, 405, 800

\bibitem[{Peletier (1993)Peletier}]{Peletier93}
Peletier, R.~F.~1993, A\&A, 271, 51

\bibitem[{Peng {et~al.}(2010)Peng, Ho, Impey, \& Rix}]{Peng:2010p4252}
Peng, C.~Y., Ho, L.~C., Impey, C.~D., \& Rix, H.-W. 2010, AJ, 139, 2097

\bibitem[{S{\'e}rsic(1963)}]{1963BAAA....6...41S}
S{\'e}rsic, J.~L. 1963, Bol. Asoc. Argentina Astron. La Plata Argentina, 6, 41

\bibitem[{Smith {et~al.}(2010)Smith, Davies, \& Nelson}]{Smith10}
Smith, R., Davis, J.~I., \& Nelson, A.~H. 2010, MNRAS, 405, 1723

\bibitem[{Smith {et~al.}(2011)Smith, Fellhauer, \& Assmann}]{Smith11}
Smith, R., Fellhauer, M., \& Assmann, P. 2011, MNRAS accepted, arXiv:1110.5555

  \bibitem[{Thompson (1981)}]{Thompson}
  Thompson, L.~A. 1981, ApJ, 244, L43
  

\bibitem[{Toloba {et~al.}(2011)Toloba, Boselli, Cenarro, Peletier, Gorgas,
  de~Paz, \& Mu{\~n}oz-Mateos}]{Toloba:2011p3986}
Toloba, E., {et~al.} 2011, A{\&}A, 526, 114

\bibitem[{Toloba {et~al.}(2009)Toloba, Boselli, Gorgas, Peletier, Cenarro,
  Gadotti, de~Paz, Pedraz, \& Yildiz}]{Toloba:2009p3937}
---. 2009, ApJ, 707, L17

\bibitem[{Weinmann {et~al.}(2011)Weinmann, Lisker, Guo, Meyer, \&
  Janz}]{Weinmann:2011p3976}
Weinmann, S.~M., Lisker, T., Guo, Q., Meyer, H.~T., \& Janz, J. 2011, MNRAS, 416, 1197

\end{thebibliography}
\end{document}